\documentclass[12pt]{iopart}
\usepackage{epsfig}    
\usepackage{inputenc}   
\usepackage{color}      
\usepackage{subfigure}  
\usepackage{amssymb}    
\usepackage{iopams}    
\begin{document}

\title{Effect of stacking faults on the magnetocrystalline anisotropy of hcp Co: a first-principles study}
\author{C.J.~Aas$^1$, L.~Szunyogh$^2$, R.F.L.~Evans$^1$, R.W.~Chantrell$^1$}
\address{$^1$ Department of Physics, University of York, York YO10 5DD, United Kingdom}
\address{$^2$ Department of Theoretical Physics and Condensed Matter Research Group of Hungarian Academy of Sciences, Budapest University of Technology and Economics, Budafoki \'ut 8.,~H1111 Budapest, Hungary}

\begin{abstract}
In terms of the fully relativistic screened Korringa-Kohn-Rostoker method we investigate the effect of stacking faults on the magnetic properties of hexagonal close-packed cobalt.  In particular, we consider the formation energy and the effect on the magnetocrystalline anisotropy energy (MAE) of four different stacking faults in hcp cobalt -- an intrinsic growth fault, an intrinsic deformation fault, an extrinsic fault and a twin-like fault.  We find that the intrinsic growth fault has the lowest formation energy, in good agreement with previous first-principles calculations.  With the exception of the intrinsic deformation fault which has a positive impact on the MAE, we find that the presence of a stacking fault generally reduces the MAE of bulk Co.  Finally, we consider a pair of intrinsic growth faults and find that their effect on the MAE is not additive, but synergic.
\end{abstract}

\maketitle
\section{Introduction}
Within the magnetic recording industry, cobalt alloys such as CoPt and CoPd are of great interest due to their large magnetocrystalline anisotropy energies (MAE) \cite{luklemmer}.  For the purpose of magnetic recording, a large MAE of the recording medium is crucial in order to maintain stability of the written information as larger areal information storage densities require smaller grain sizes \cite{weller}.  In close-packed metals and alloys, stacking faults are known to form relatively easily \cite{chetty}.  This is one of the contributing factors to the relatively large ductility and malleability that are observed in many such materials \cite{chetty}.  For a magnetic recording medium, the presence of stacking faults is generally considered detrimental,  as disturbances in the microstructure will generally worsen the signal-to-noise ratio of the medium \cite{sokalski}.  Stacking faults may also break the local lattice symmetry and, therefore, drastically influence the MAE.\\*

Experimentally, the effects of stacking faults are generally measured in terms of the stacking fault density, which can be determined from X-ray diffraction spectra (see e.g.~\cite{luklemmer,sokalski,mitra}).  There are a large number of experimental studies into stacking fault formation energies and the effect of the stacking fault density on the magnetocrystalline anisotropy for various magnetic recording alloys \cite{luklemmer,saito}.  However, in experiment, the real effect of a stacking fault might be obscured by other phenomena, such as migration of impurities along the stacking fault, synergies of closely spaced stacking faults, etc.  Consequently, a number of theoretical methods have been developed for determining the properties and effects of stacking faults, see e.g.~\cite{berliner}.  In particular, there is a large number of first-principles studies of the formation energies of given types of stacking faults in metals~\cite{chetty, crampin}.  It has been suggested that stacking fault formation energies determined from first-principles may be more accurate than experimental measurements \cite{chetty} as theoretical calculations separate the formation energy from any other correlated effects on the total energy.  The effect on the MAE of a particular stacking fault is, however, less commonly explored.  In this work, we aim to determine from first principles the effect on the MAE of four different types of stacking faults in hcp cobalt.

\section{The stacking faults}
\label{sec:stackingfaults}
Hexagonal planes can be packed either in an ...ABAB... sequence, yielding a hexagonal close-packed lattice structure, or in an ...ABCABC... sequence, yielding a face-centred cubic lattice structure \cite{physmetbook}.  In the hexagonal close-packed lattice structure, the stacking direction corresponds to the $(0 0 0 1)$ axis of the lattice, whereas for the face-centred cubic lattice structure, the stacking direction is parallel to the $(1 1 1)$ axis of the lattice.  In a hcp lattice, a stacking fault is defined as an interruption in the ...ABAB... stacking of the hexagonal planes.  While there are of course any number of conceivable stacking faults, their varying degrees of formation energies and formation mechanisms mean they have different probabilities of occurrence \cite{crampin}.  In line with previous work \cite{chetty,zope}, we consider the following four different stacking faults, denoted in standard notation as I$_1$, I$_2$, E and T$_2$ \cite{sfbook1, frank}:
\begin{itemize}
 \item I$_1$ (intrinsic): $\cdots$ B  A  B  A  ${\mathrm{\bf B}}$  C  B  C  B  $\cdots$
 \item I$_2$ (intrinsic): $\cdots$ A  B  A  B ${\bf \mid}$ C  A  C  A  $\cdots$
 \item E (extrinsic): $\cdots$ A  B  A  B  ${\mathrm{\bf C}}$  A  B  A  B  $\cdots$
 \item T$_2$ (twin-like): $\cdots$ A  B  A  B  ${\mathrm{\bf C}}$  B  A  B  A  $\cdots$
\end{itemize}
Here the bold face letters or vertical line denote the plane of reflection symmetry of the stacking fault.  In an \emph{intrinsic} stacking fault (I$_1$ and I$_2$), the stacking fault is simply a shift of one lattice parameter and the stacking on either side is correct all the way up to the very fault \cite{physmetbook}.  The stacking fault I$_1$ is a growth fault while the stacking fault I$_2$ is a deformation fault \cite{chetty}.  In the \emph{extrinsic} stacking fault (E), a plane has been inserted so that it is incorrectly stacked with respect to the planes on either side of it \cite{physmetbook, Hammer}.  In a \emph{twin}-like fault (T$_2$), the stacking sequence is reflected in the fault layer \cite{chetty}.  In the following, we refer to the centre of reflection symmetry as the zeroth layer.  The two layers adjacent to the centre of reflection symmetry are then indexed $\pm 1$, and so on.  Note that in the case of a stacking fault of type I$_2$, the plane of reflection symmetry lies in between two atomic layers.  Therefore, in the following, for type I$_2$ the atomic layers will be labelled by $\pm \frac{1}{2},  \pm \frac{3}{2}, \dots$, rather than by $0, \pm 1, \pm 2, \dots$ as for the other types of stacking faults.

\section{Computational details}
For our theoretical study we employed the fully relativistic Screened Korringa-Kohn-Rostoker (SKKR) method, in which the Kohn-Sham scheme is performed in terms of the Green's function of the system (rather than the wavefunctions) and the treatment of extended layered systems is particularly efficient~\cite{laszlo1,screening1,screening2}.  We used the local spin density approximation (LSDA) of density functional theory (DFT) as parametrised by Vosko and co-workers~\cite{voskoCJP80}.  The effective potentials and fields were treated within the atomic sphere approximation (ASA) and an angular momentum cut-off of $\ell_{max}=2$ was used.   The magnetocrystalline anisotropy energy (MAE) is calculated using the magnetic force theorem~\cite{jansen}, within which the MAE is defined as the difference in the band energy of the system when magnetised along the easy axis $( 0 0 0 1)$ and perpendicular to the easy axis. Alternatively, the uniaxial MAE can be calculated from the derivative of the band energy, for more details see Ref.~\cite{aas-PtCo}.  Only when calculating the MAE, we used an angular momentum cut-off of $\ell_{max}=3$. \\*

The LSDA+ASA fails in describing the orbital moment and the MAE of Co accurately.  Similar to our previous work~\cite{aas-PtCo} we, therefore, employed  the orbital polarisation (OP) correction \cite{orbcorr, eschrig, eschrig2}, as implemented within the KKR method by Ebert and Battocletti \cite{orbcorrKKR}.  Note that the OP correction was applied only for the $\ell=2$ orbitals.  Excluding the OP correction we obtained an easy-plane magnetisation and a MAE of 6.7 $\mu$eV per cobalt atom, while including the OP correction we obtained an easy axis perpendicular to the hexagonal cobalt planes and a MAE of 84.4 $\mu$eV per cobalt atom.  This is in good  agreement with the experimental value of 65.5 $\mu$eV \cite{stearns} and with the experimental easy axis being parallel to the $(0 0 0 1)$ direction. Our result also compares well with that of Trygg {\em et al.} \cite{trygg95}, who calculated $K=110$~$\mu$eV for hcp cobalt using a full-potential LMTO method including the OP correction.\\*

In this study we consider an infinite cobalt system, consisting of two semi-infinite bulk cobalt systems and an internal region (region $I$).  Region $I$ contains the stacking fault and is positioned in between the semi-infinite regions.  The combined system is periodic and infinite in the plane normal to the $(0 0 0 1)$ direction.  Due to the long-ranged nature of the stacking fault effects on the MAE (see section \ref{sec:aggMAECo}), the region $I$ in this study had to be kept at a size of around 80 atomic layers. More specifically, for stacking faults I$_1$ and I$_2$, systems of 80 atomic layers were required, while for stacking faults E and T$_2$, 74 atomic layers were required. In order to keep the calculations tractable we limited the self-consistent calculations only  for a number of atomic layers near the stacking fault, and then appended the bulk potentials for the atomic layers further away from the stacking fault.  We found that it was sufficient to treat only the 20 centremost layers self-consistently.  In line with previous first-principles studies of stacking faults in close-packed metals, we ignored any atomic and volume relaxations (see e.g.~\cite{crampin}).  The effects of such relaxations are normally negligible because atoms in the faulted part of the system tend to retain their close-packed coordination numbers despite the presence of the fault \cite{chetty, Hammer, denteneer, twins, schweizer, wrightSF}.  Throughout, therefore, we have used the experimental lattice parameter for cobalt, $a=2.507$ \AA.

\section{Results}
\subsection{Stacking Fault Formation Energies}
Before exploring how the stacking faults influence the  MAE of bulk Co, we would like to gain an idea of their formation energy. Within the SKKR-ASA scheme, the LSDA total energy can be cast into contributions related to individual atomic cells, $E_i$, comprising the kinetic energy, the intracell Hartree energy and the exchange-correlation energy, and into the two-cell Madelung (or intercell Hartree) energy, $E_{\rm Mad}$~\cite{laszlo1}.  For a simple bulk metal, like hcp Co, $E_{\rm Mad}$ is, in practice, negligible, while in the presence of stacking faults it gives a non-negligible contribution due to charge redistributions.  However, from our self-consistent calculations we found that $E_{\rm Mad}$  is in the order of $0.1-0.2$ meV per stacking fault.  Since the typical formation energy of stacking faults are
by about two orders higher in magnitude than this value, in the following we consider only the layer-resolved (cell-like) contributions to the  the total energy.  In order to check these contributions for artefacts of the appending of the bulk potential, we compare $E_i$  for $i=-10$ (layer with effective potential from a self-consistent stacking  fault calculation) to $E_i$  for $i=10$ (layer with appended bulk potential), since due to the mirror symmetry, these two contributions should be identical. Reassuringly enough, they agreed to within a relative error of $10^{-9}$, which is well within intrinsic and numerical error of our computational method.\\*

\begin{figure}[htp!]
\begin{center}
\fontsize{12}{12}\selectfont
\includegraphics[scale=0.4]{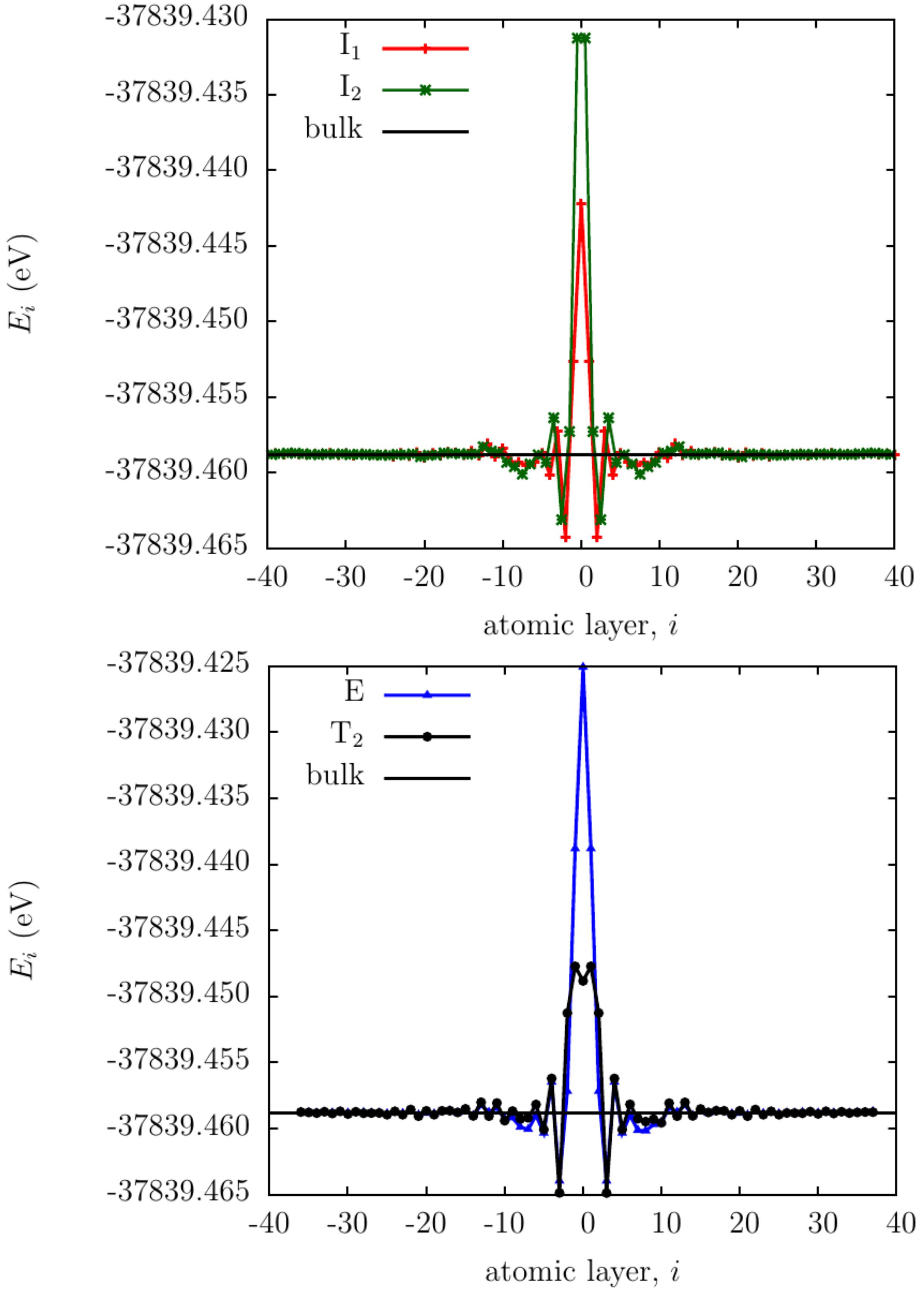}
\caption{The layer-resolved contributions to the total energy in four hcp cobalt systems, each exhibiting one of the four different types of stacking fault. The label $0$ refers to the plane of mirror symmetry. Solid lines serve as a guide for the eyes.
\label{fig:layer-totEres}}
\end{center}
\end{figure}

The layer-resolved contributions to the total energy across the systems containing the stacking faults I$_1$, I$_2$, E and T$_2$ is shown in Fig.~\ref{fig:layer-totEres}.  Herein we observe the expected mirror symmetry and that the layer-resolved total energy contributions approach the bulk total energy, $E_{Co}=-37839.459$ eV, towards the edges of each system. From this figure it is obvious that the deviation of $E_i$  from $E_{Co}$ is significant up to about 15 layers away from the centre of stacking fault. \\*

The stacking fault formation energy is defined as the difference in the total energy caused by the presence of the stacking fault. In order to ensure we include enough atomic layers on either side of the stacking fault, we consider the cumulative sums:
 \begin{equation}
\Delta E_{(\mathrm{I}_1, \mathrm{E},\mathrm{T}_2)}(N) =
\sum_{-N}^{N}E_{i}  - (2N+1) E_{Co}  \: , \label{eq:Esum1}
\end{equation}
and
 \begin{equation}
\Delta E_{\mathrm{I}_2}(N) = \sum_{-N+\frac{1}{2}}^{N-\frac{1}{2}}E_{i} - 2N E_{Co} \: .
\label{eq:Esum2}
\end{equation}
The formation energy of a given stacking fault $X=\mathrm{I}_1,\mathrm{I}_2, \mathrm{E},\mathrm{T}_2$, $E_{form}^{(X)}$, is then defined as
\begin{equation}
 E_{form}^{(X)} = \lim_{N \rightarrow \infty} \left( \Delta E_X (N) \right) \: .
\end{equation}

\begin{figure}[htp!]
\begin{center}
\fontsize{12}{12}\selectfont
\includegraphics[scale=0.4]{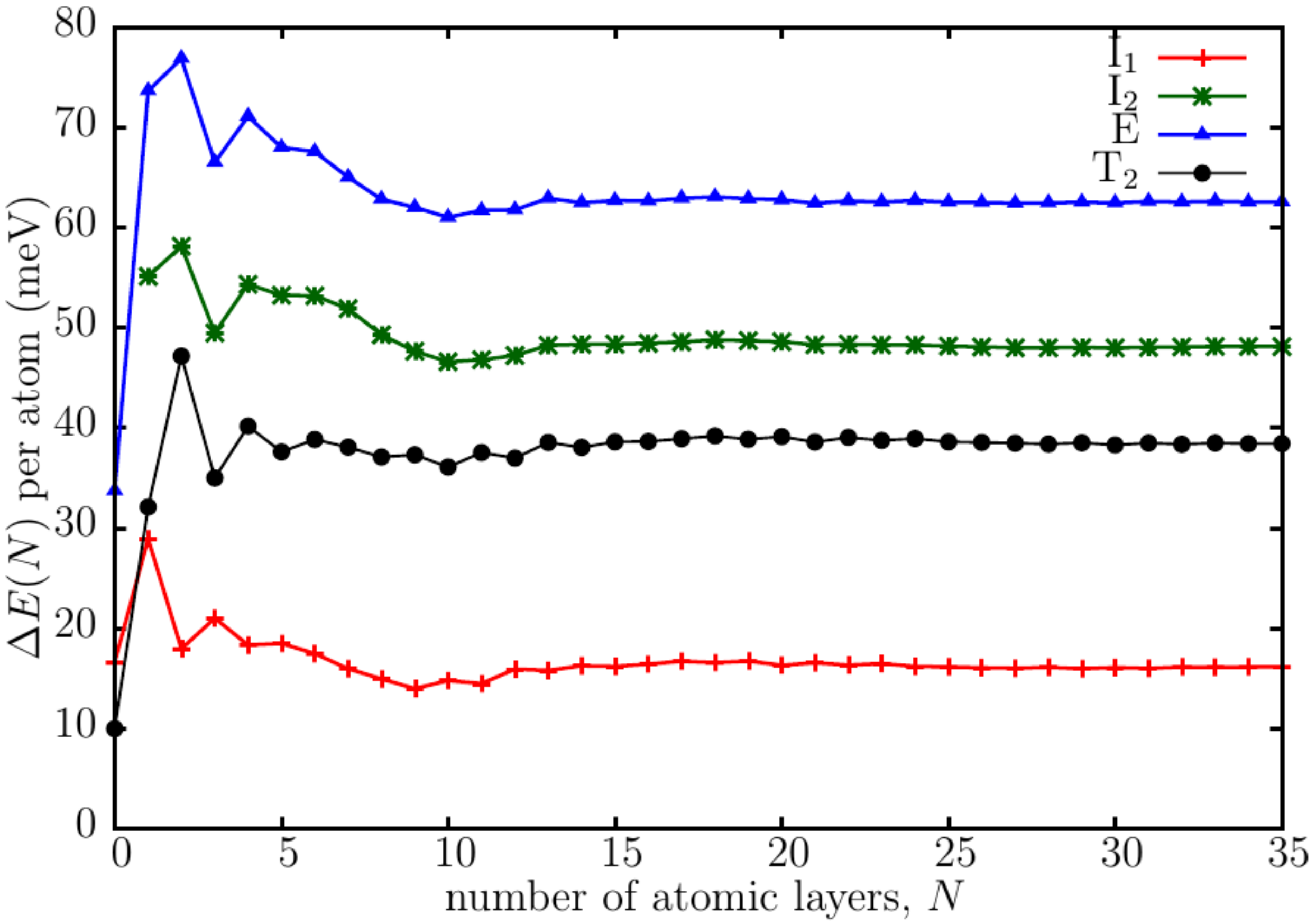}
\caption{The cell-like part of the formation energy, $\Delta E_X (N)$, see Eqs.~(\ref{eq:Esum1}) and (\ref{eq:Esum2}),  of  stacking faults  I$_1$, I$_2$, E and T$_2$ in hcp cobalt, displayed as a function of the number of layers, $N$, considered in the system on either side of the stacking fault.  Solid lines serve as guides for the eyes.
\label{fig:lmax2totE}}
\end{center}
\end{figure}

The calculated values of $\Delta E_{X}(N)$ are shown in Fig.~\ref{fig:lmax2totE}.  Quite obviously, for all types of stacking faults, nearly 30 atomic layers (i.e., 15 layers on either side of the stacking fault) are required in order to obtain well-converged stacking fault formation energies.  The fact that the effect of the stacking fault is relatively long-ranged could have significant impact on nano-sized systems as the formation energy and, consequently, the likelihood of occurrence of a stacking fault could be different depending on its location in relation to, e.g., other imperfections as well as surfaces or interfaces in the sample.  We obtain the following formation energies, with a possible error of $\sim 0.1-0.2$ meV due to the Madelung energy not being included:
\begin{eqnarray}
E_{form}^{(I_1)} & \approx & 16 \mathrm{\ meV \ } \approx 40 \mathrm{\ mJ}\cdot\mathrm{m}^{-2} \nonumber \\
E_{form}^{(I_2)} & \approx & 48 \mathrm{\ meV \ } \approx 122 \mathrm{\ mJ}\cdot\mathrm{m}^{-2} \nonumber  \\
E_{form}^{(E)}   & \approx & 62 \mathrm{\ meV \ } \approx 160 \mathrm{\ mJ}\cdot\mathrm{m}^{-2} \nonumber  \\
E_{form}^{(T_2)} & \approx & 39 \mathrm{\ meV \ } \approx 100 \mathrm{\ mJ}\cdot\mathrm{m}^{-2} \: . \nonumber
\end{eqnarray}
As expected, all stacking faults incur a positive change in the total energy.   Of the four types of stacking faults considered here, the intrinsic stacking fault I$_1$ has the lowest formation energy  and  the stacking fault E exhibits the highest one.  While there is no available experiment in literature, the overall results agree well with e.g. Ref.~\cite{crampin}: the extrinsic stacking fault formation energy for the close-packed fcc metals in this study is generally significantly larger than that of the intrinsic and twin faults. Moreover, our calculated values for the hcp Co growth stacking fault I$_1$ and the hcp Co extrinsic fault E are close to those obtained by Crampin and co-workers for Ni (which is next to Co in the periodic table)~\cite{crampin}: 28 $\mathrm{ mJ}\cdot\mathrm{m}^{-2}$ for the intrinsic stacking fault  and 180 $\mathrm{ mJ}\cdot\mathrm{m}^{-2}$ for the extrinsic fault.

\subsection{Layer-Resolved Contributions to the Magnetocrystalline Anisotropy Energy}
\begin{figure}[htp!]
\begin{center}
\fontsize{12}{12}\selectfont
\includegraphics[scale=0.4]{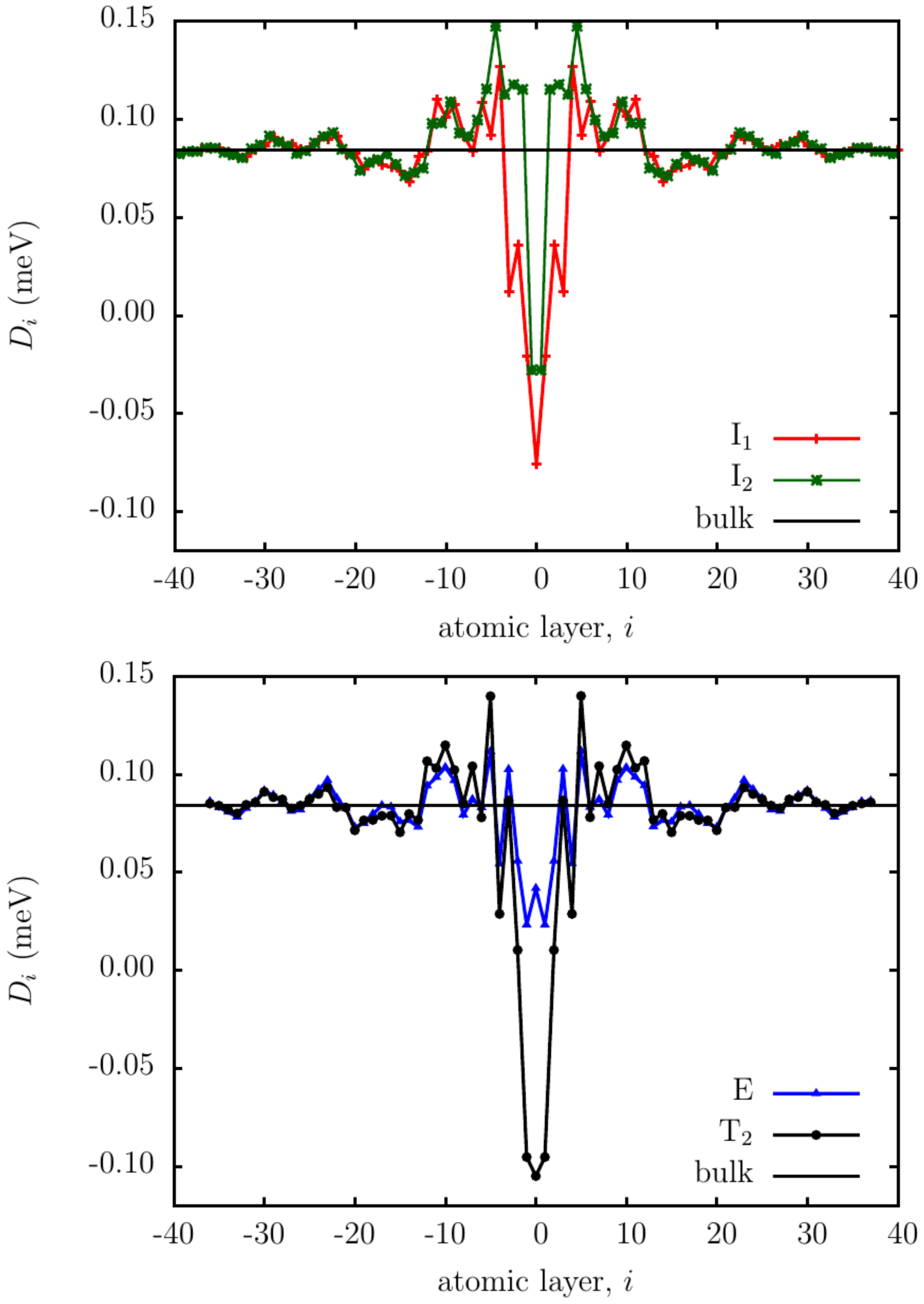}
\caption{Calculated layer-resolved contributions to the MAE for stacking faults I$_1$ and $I_2$ (upper panel) and E and T$_2$ (lower panel).  The horizontal line refers to the bulk MAE, 84.4 $\mu$ eV/atom.
Solid lines serve as guide for the eyes.
\label{fig:layer-res}}
\end{center}
\end{figure}

Because it is calculated directly from the band energy, the MAE can naturally be resolved into layer-dependent contributions, $D_i$, see Ref.~\cite{aas-PtCo}.  These layer-resolved contributions are depicted in Fig.~\ref{fig:layer-res} for the different types of stacking faults. Note that the mirror symmetry is well reproduced in the layer-resolved MAE contributions for all stacking faults.  Moreover, the MAE approaches the bulk MAE, $K_{Co}=84.4$ $\mu$eV, towards the edges of all four systems.  For stacking faults of type I$_1$, I$_2$ and T$_2$, the MAE contributions become negative at the centre of the fault, favoring thus an in-plane easy axis in these layers.  This could indicate that these types of stacking faults may act as pinning sites.  For the type E stacking fault, the layer-resolved MAE contributions near the centre are also reduced, retaining, however, very small positive values.\\*

Furthermore, we note that all stacking faults induce long-ranged oscillations in the MAE.   For layers of about $\left| i \right| > 15$, the four stacking faults exhibit very similar trends in the layer-resolved MAE contributions.  In other words, at about 15 layers away from the stacking fault, the presence of a stacking fault still influences the MAE, while the particular type of the stacking fault is less significant.  This will, however, obviously depend on the size of the sample.

\subsection{Finite Size Effect on the Magnetocrystalline Anisotropy Energy}
\label{sec:aggMAECo}
The long-ranged oscillations in the MAE could cause significant finite-size effects in the experimental determination of the MAE of nano-sized samples.  We therefore consider the following cumulative sums,
\begin{equation}
K_{(\mathrm{I}_1, \mathrm{E},\mathrm{T}_2)}(N) =
\sum_{-N}^{N}D_{i}  - (2N+1) K_{Co}  \: , \label{eq:Ksum1} 
\end{equation}
and
\begin{equation}
K_{\mathrm{I}_2}(N) =  \sum_{-N+\frac{1}{2}}^{N-\frac{1}{2}}D_{i} - 2N K_{Co} \: ,
\label{eq:Ksum2}
\end{equation}
where the MAE of the stacking fault systems of finite width is related to the MAE of hcp Co. \\*

\begin{figure}[htp!]
\centering
\fontsize{12}{12}\selectfont
\includegraphics[scale=0.4]{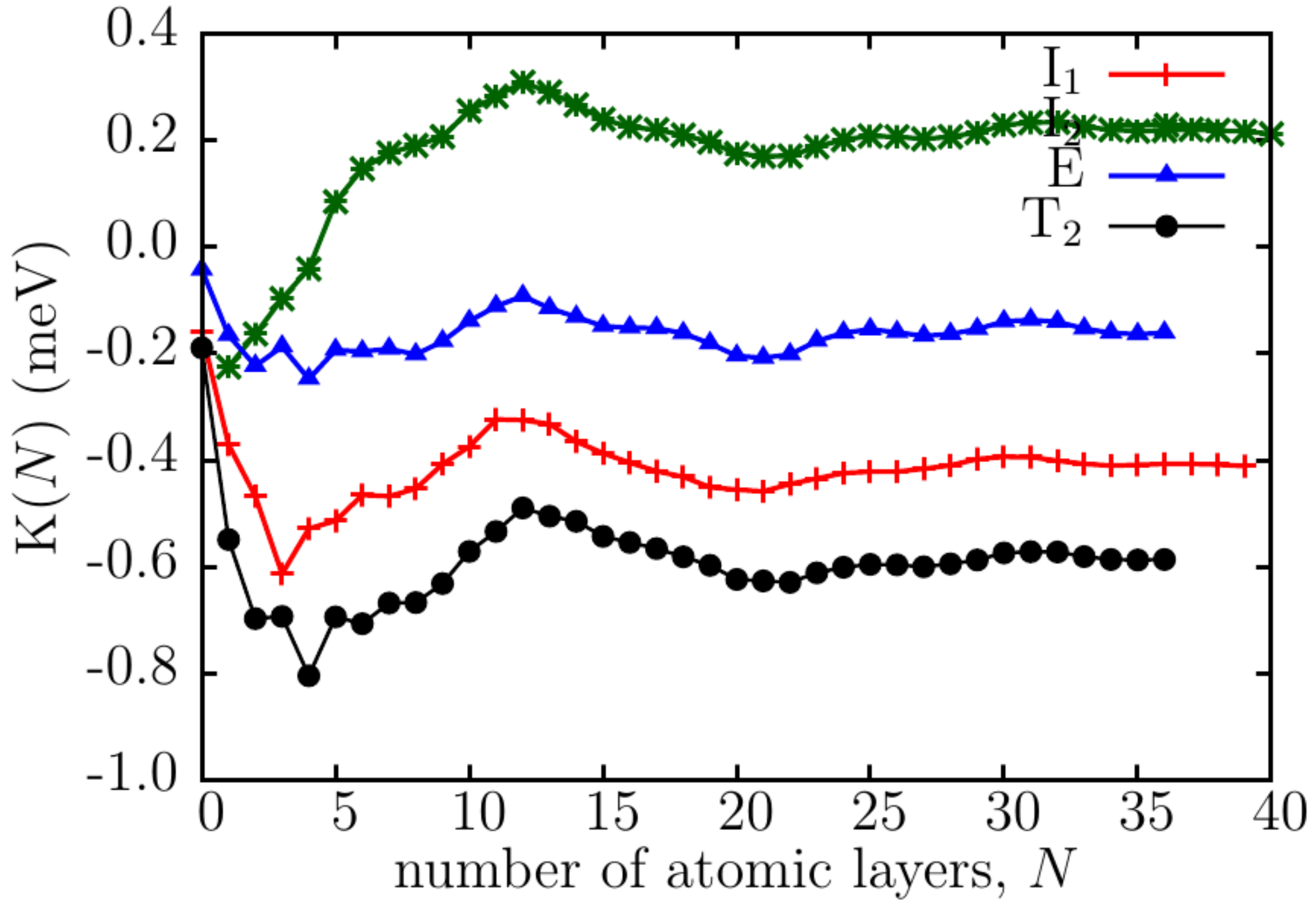}
\caption{Cumulative sums of layer-resolved contributions to the MAE, $K_{X}(N)$  ($X=$  I$_1$, I$_2$, E and T$_2$), see Eqs.~(\ref{eq:Ksum1}) and (\ref{eq:Ksum2}).
Solid lines serve as guide for the eyes.  \label{fig:friedel-tot}}
\end{figure}

Fig.~\ref{fig:friedel-tot} shows $K_{X}(N)$ for the four different stacking faults as a function of $N$.  Surprisingly, for $N \ge 5$ the I$_2$ type stacking fault appears to increase the MAE, i.e., to strengthen the easy axis $(0 0 0 1)$  ({\em positive effect}).  As seen from Fig.~\ref{fig:layer-res}, this is due to the positive MAE contributions induced by the stacking fault on neighbouring layers.  These apparently outweigh the strongly negative MAE contributions induced in the centre of stacking fault type I$_2$.  This is an unexpected result as stacking faults are typically reported to lower the MAE (see e.g.~\cite{sokalski}).  It should be noted, however, that, of the stacking faults studied here, type I$_2$ has the next highest formation energy and it is therefore less likely to occur in an equilibrated sample.  For stacking faults of types I$_1$, E and T$_2$, the overall change in the MAE with respect to hcp Co is negative ({\em negative effect}). As noted earlier, in the vicinity of these stacking faults, the easy dirction is rotated normal to the $(0001)$ axis.  This is consistent with the reduction in the total MAE observed experimentally by Sokalski et al.~in\cite{sokalski}.\\* 

It is quite a remarkable feature that, as seen from Fig.~\ref{fig:friedel-tot}, the layer-resolved MAE contributions do not settle until at about approximately 35 layers on either side of the stacking fault.   This long-ranged behaviour could give rise to significant finite-size effects in nano-sized samples.  Moreover, this might have consequences for theoretical investigations into the formation and effects of stacking faults on magnetic properties.  Typically, in Monte Carlo simulations of stacking faults, interactions between stacking faults is kept to around three neighbouring planes \cite{sokalski}.  In light of our results, this appears to be an uncertain assumption.

\subsection{Magnetocrystalline Anisosotropy of a Composite Stacking Fault}
Experimentally, the presence of stacking faults is normally quantified in terms of the stacking fault density, which is partly a measure of how close the stacking faults are located.  As the simplest assumption, the change in the MAE due to the presence of a number stacking faults in a sample is approximated by the sum of the changes in the MAE due to each individual stacking fault.  If this were the case, the effect of an isolated stacking fault on the MAE could quite straightforwardly be transformed into the change in MAE as a function of the stacking fault density.  However, the long-ranged oscillations in the MAE caused by the presence of each stacking fault indicates that the situation is far more complex.\\*

In particular, we considered two stacking faults of type I$_1$, separated by three atomic layers.  In other words, the system exhibits the composite stacking fault:
\begin{center}
$\cdots$   A  B  A  ${\mathrm{\bf B}}$  C  B  C  ${\mathrm{\bf B}}$ A  B  A  $\cdots$
\end{center}
Note that by removing one of the two C-B pairs of atomic layers, a twin-like stacking fault T$_2$ is obtained. We have chosen three atomic layers between the centres of the two stacking faults, since in dynamical models it is often used as the distance beyond which the interaction between stacking faults is neglected (see e.g. Ref.~\cite{sokalski}).  Moreover, we deal with a pair of  I$_1$ type stacking faults  because this type of stacking fault has the lowest formation energy and is, therefore, expected to occur more commonly than the other three types of stacking faults.\\*

\begin{figure}[htp!]
\fontsize{12}{12}\selectfont
\centering
\includegraphics[scale=0.4]{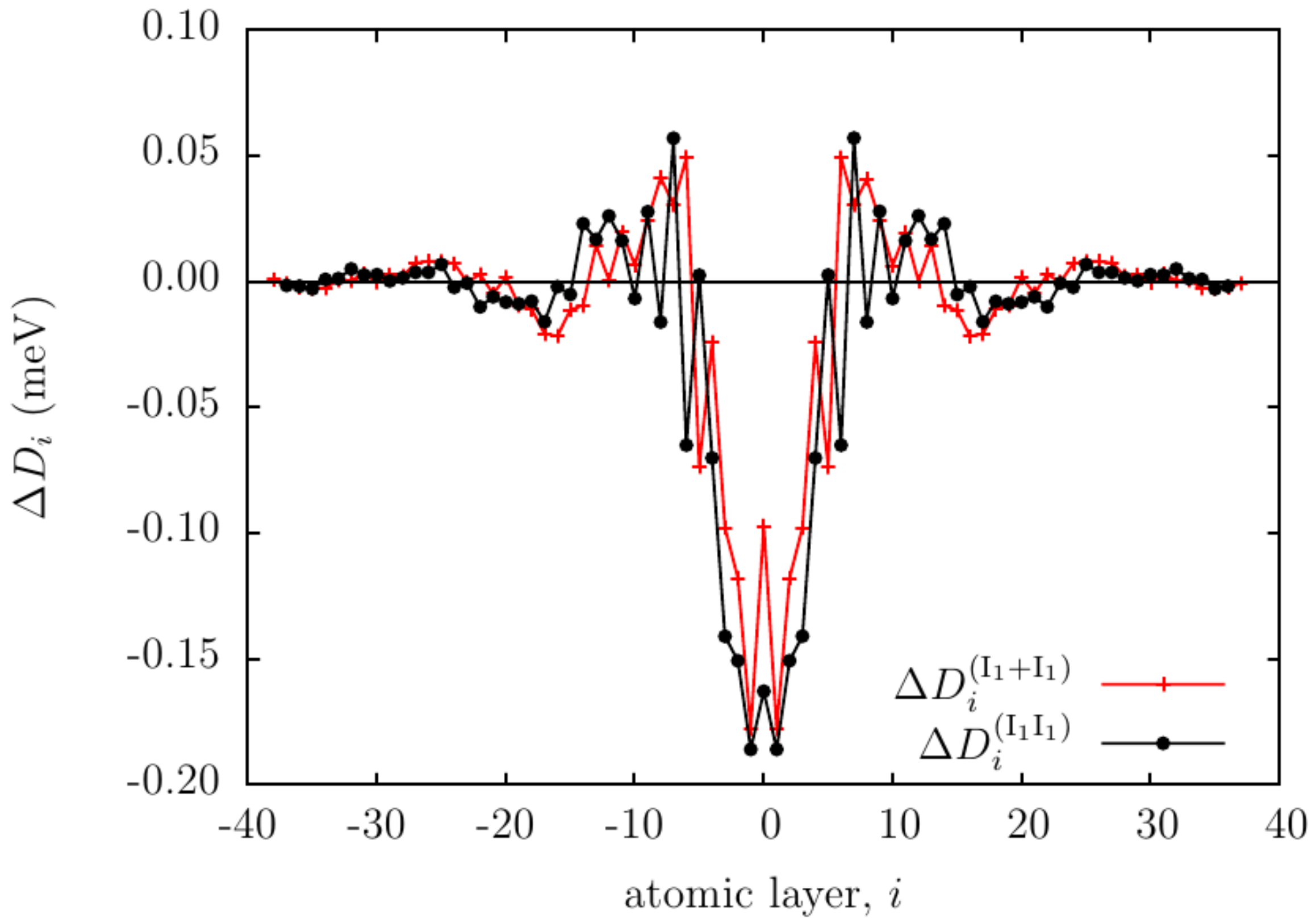}
\caption{$\bullet:$ Calculated deviations in the layer-resolved MAE contributions, $\Delta D_i^{(\mathrm{I}_1 \mathrm{I}_1)}$, of the composite stacking fault from the bulk Co MAE, see Eq.~(\ref{eq:DeltaDII}), and $+ :$
the corresponding average deviations, $\Delta D_i^{(\mathrm{I}_1+\mathrm{I}_1)}$, of two superposed I$_1$ type stacking faults centred on atomic layers $i \pm 2$, see Eq.~(\ref{eq:DeltaDI-I}). Solid lines serve as guide for the eyes. \label{fig:averagecomp}}
\end{figure}

The difference between the layer-resolved MAE contributions and the MAE of bulk hcp Co,
\begin{equation}
\Delta D_i^{(\mathrm{I}_1\mathrm{I}_1)} =  D_{i}^{(\mathrm{I}_1\mathrm{I}_1)} - K_{Co} \: ,
\label{eq:DeltaDII}
\end{equation}
is shown in Fig.~\ref{fig:averagecomp} for the composite stacking fault.  As a comparison, we also show the average deviations in the layer-resolved MAE contributions from the bulk MAE of two independent type I$_1$ stacking faults,
\begin{equation}
\Delta D_i^{(\mathrm{I}_1+\mathrm{I}_1)} = \frac{1}{2} \left(D_{i+2}^{(\mathrm{I}_1)} + D_{i-2}^{(\mathrm{I}_1)}\right)  - K_{Co} \: .
\label{eq:DeltaDI-I}
\end{equation}
If $\Delta D_i^{(\mathrm{I}_1\mathrm{I}_1)}$ and $\Delta D_i^{(\mathrm{I}_1+\mathrm{I}_1)}$ were equal for each atomic layer $i$, the change of the MAE due to the presence of the composite stacking fault would be exactly twice that of a single I$_1$ stacking fault.  However, as shown in Fig.~\ref{fig:averagecomp}, $\Delta D_i^{(\mathrm{I}_1\mathrm{I}_1)}$ and $\Delta D_i^{(\mathrm{I}_1+\mathrm{I}_1)}$ deviate significantly, particularly in the layers $\left|i\right|  \le 2$, i.e., in the layers between the two stacking faults.  Beyond $\left|i\right| >3$, the magnitudes of the MAE contributions are similar for  $\Delta D_i^{(\mathrm{I}_1\mathrm{I}_1)}$ and $\Delta D_i^{(\mathrm{I}_1+\mathrm{I}_1)}$, but with a phase shift of approximately one layer.\\*

\begin{figure}[htp!]
\centering
\fontsize{12}{12}\selectfont
\includegraphics[scale=0.4]{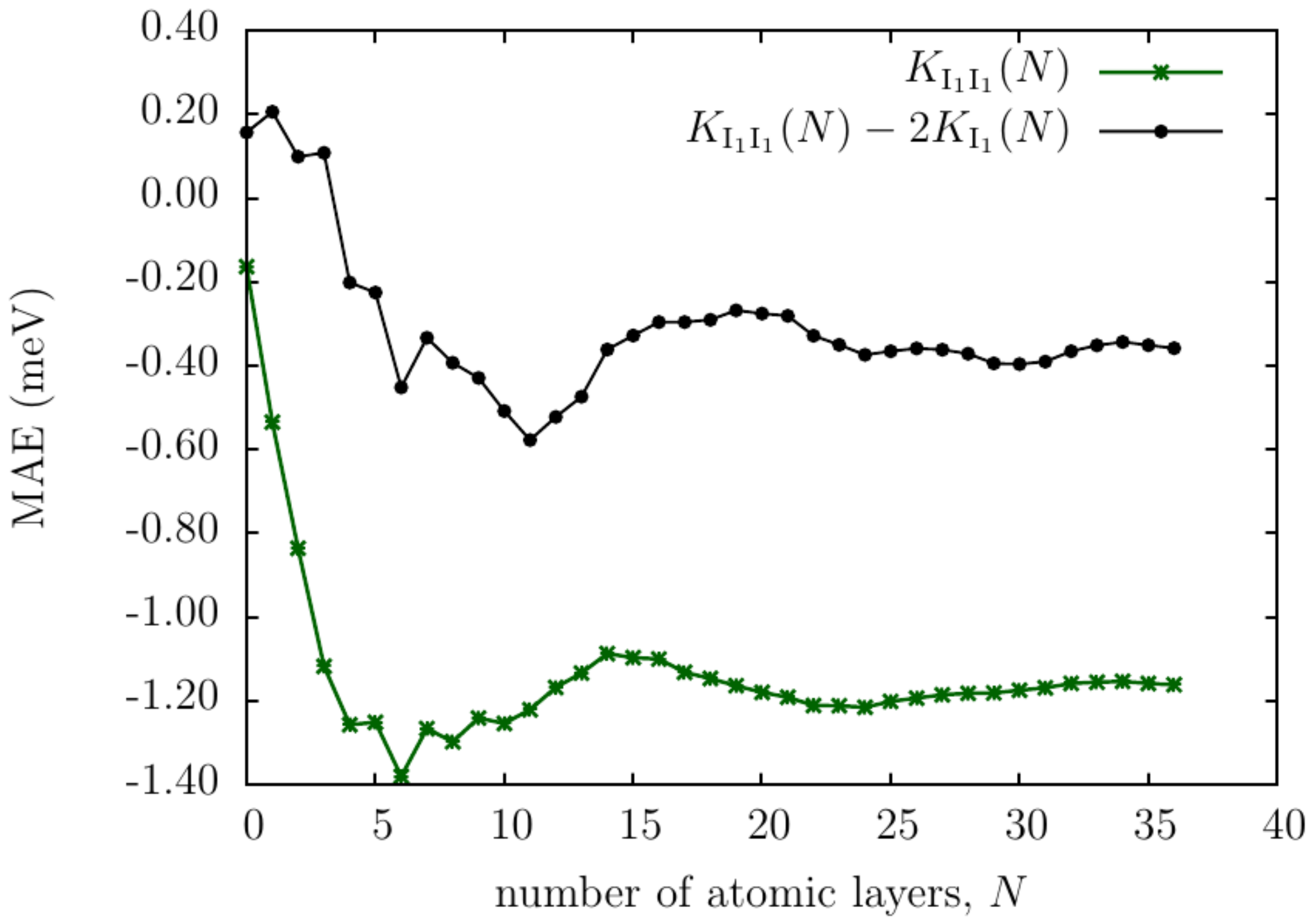}
\caption{$\star:$ Change in the MAE of hcp Co due the composite stacking fault, $K_{\mathrm{I}_1 \mathrm{I}_1}(N)$, as defined in Eq.~\ref{eq:compaggMAE}).  $\bullet :$ Interaction term of the two stacking faults in the MAE, $K_{I_1 I_1}(N) - 2K_{I_1}(N)$, see.~Eqs.~(\ref{eq:Ksum1}) and (\ref{eq:compaggMAE})
Solid lines serve as guide for the eyes.
.  \label{fig:KiiM2Ki}}
\end{figure}

Similar to the case of single stacking faults, we calculate the cumulative sum of the MAE contributions of the composite stacking fault,
\begin{equation}
 K_{\mathrm{I}_1 \mathrm{I}_1}(N) = \sum_{i=-N}^{N} D_i^{(\mathrm{I}_1 \mathrm{I}_1)} - (2N+1) K_{Co}
 \: ,
 \label{eq:compaggMAE}
\end{equation}
and plot it in Fig.~\ref{fig:KiiM2Ki}. Apparently, $K_{\mathrm{I}_1 \mathrm{I}_1}(N)$ converges to approximately $-1.18$ meV for large $N$, which is almost three times the change of the MAE of the single type I$_1$ stacking fault ($\sim -0.40$ meV, see Fig.~\ref{fig:friedel-tot}). Also shown in Fig.~\ref{fig:KiiM2Ki} is the difference $K_{\mathrm{I}_1 \mathrm{I}_1}(N) -2 K_{\mathrm{I}_1}(N)$, which appears to settle at approximately $-0.38$ meV.  In other words, the two stacking faults interact to yield a stronger negative effect on the total MAE as compared to two isolated type I$_1$ stacking faults.  This appears to be mainly due to MAE contributions from the atomic layers located in between the two type I$_1$ stacking faults.  This could have significant consequences for predicting the resulting MAE in dynamical models used to explain experimental data.  To draw any definite conclusions, a systematic study of the stacking fault types and separations would be required.  We expect that such a study would be computationally extremely intensive as interlayers (or supercells) of up to approximately 160 atomic layers would be required in order to reach the limit in which the two stacking faults are far enough apart not to interact.

\section{Summary and Conclusions}
Using the fully relativistic screened Korringa-Kohn-Rostoker method, we have studied the MAE of bulk hcp cobalt in the vicinity of four different types of stacking faults.  We find that, in accordance with experiment, most stacking faults have a detrimental overall effect on the MAE.  The one exception to this overall conclusion is the type $I_2$ intrinsic stacking fault, which, however, exhibits a relatively high formation energy and which may, consequently, occur relatively infrequently under standard experimental conditions.  The effect of a stacking fault on the layer-resolved contributions to the MAE is long-ranged, in the order of 15 atomic layers on either side of each stacking fault.  Motivated by this observation, we investigated a particular composite stacking fault and concluded that the MAE of the composite stacking fault is not identical to the sum of the MAE of the two isolated stacking faults. A further challenging study is proposed regarding the dependence of the 'interaction' of two stacking faults on the separation between them.\\*

CJA is grateful to EPSRC and to Seagate Technology for the provision of a research studentship.
Support of the EU under FP7 contract NMP3-SL-2012-281043 FEMTOSPIN is gratefully acknowledged.
Financial support was in part provided by the New Sz\'echenyi Plan of Hungary (T\'AMOP-4.2.2.B-10/1--2010-0009)
and the Hungarian Scientific Research Fund (OTKA K77771). \\*

\end{document}